# Improved Online Wilson Score Interval Method for Community Answer Quality Ranking

Xin Cao
Georgia Institute of Technology
xincao@gatech.edu

*Abstract*: **In this paper, a fast and easy-to-deploy method with a strong interpretability for community answer quality ranking is proposed. This method is improved based on the Wilson score interval method [Wilson, 1927], which retains its advantages and simultaneously improve the degree of satisfaction with the ranking of the high-quality answers. The improved answer quality score considers both Wilson score interval and the spotlight index, the latter of which will be introduced in the article. This method could significantly improve the ranking of the best answers with high attention in diverse scenarios.**

## INTRODUCTION

The core algorithm of the Questions and Answers (Q&A) community is the ranking of the answer quality. The factors affecting the quality of answers include but not limited to: the number of up-vote and down-vote, the posting time of answers, the weighted power or influence of an answerer and voter in a specific field, e.g., if a software engineer has given many high-quality answers in the field of software development, his future answers or vote in that field could be given higher weight. Recent years, famous communities with hundreds of millions of users start to use Wilson score interval method as their core algorithm for the answer quality ranking, such as Reddit [Salihefendic, 2015], Zhihu (Chinese version of Quora). The Wilson's method modified a certain amount of unreliability when the sample size (up-vote and down-vote numbers) is small compared to the normal approximation interval and performs very well in the ranking decision of the high-quality answers. It can be also used in other context such as the voting for the customer reviews (e.g. vote for Helpful or Not Helpful) in the electronic commerce companies such as Amazon, eBay, JingDong (Chinese version of Amazon), TaoBao (Chinese version of eBay), or the up-vote or down-vote for the video comments in the YouTube. The introduction of the Wilson score interval method would improve the shopping experience of the customers and the watching experience of the audiences.

$$\begin{array}{c} Wilson\ Score\ Interval:\\ (with\ a\ fixed\ z_{1-\frac{\alpha}{2}})\\[4pt] n = u + d\\ p = u/n\\[4pt] W(p,n) = \dfrac{p + \dfrac{z_{1-\frac{\alpha}{2}}^2}{2n} \pm \dfrac{z_{1-\frac{\alpha}{2}}}{2n}\sqrt{4n(1-p)p + z_{1-\frac{\alpha}{2}}^2}}{(1 + \dfrac{z_{1-\frac{\alpha}{2}}^2}{n})} \end{array}$$

Wilson score interval method is shown as the equations above, $u$ and $d$ are the numbers of up-vote and down-vote, $n$ is the sum of $u$ and $d$, $p$ is the ratio between $u$ and $n$, $z_{1-\frac{\alpha}{2}}$ is the confidence interval parameter which is used to identify the confidence interval with the confidence level of $1-\alpha$. The lower and upper bound of Wilson Score Interval are given as $W$ in the equations. In practice, scoring the community answers for ranking usually uses the lower bound for the security reason or the conservative viewpoint.

However, there are a few problems in practice which were not covered by the Wilson Score Interval method. For instance, it suppresses the ranking of those highly controversial answers. Many questions with potential distinct viewpoints have not a unique exact "correct" answer, and may receive a large amount of but almost equal numbers of up-vote and down-vote. These answers are usually helpful and enlightening for users or visitors but only receive low ranking scores by the Wilson's method since the method focuses mainly on the up-vote ratio of the answers instead of the absolute number of voting, the latter of which though is important in many types of questions.

Another problem is that the excessive focus on the up-vote ratio may induce answerers to give more neutral-style but less enlightening answers in order to receive higher ranking in some types of questions, which has a reverse or negative impact on the community, although it might be improved by a certain amount through adjusting the confidence interval parameter $z_{1-\alpha/2}$.

Therefore, some algorithm considers many more factors besides the number of up-vote and down-vote in order to further improve the users' experience about the answer quality ranking. However, the more complicated factors might not be directly related to a specific answer as the number of up-vote and down-vote do, which frequently bring higher uncertainty or instability for a good ranking. Some studies tried to use deep learning model to predict a potential

**First Draft was Completed in April 2018**

good ranking for the answer quality, which performs well in some situations especially in the prediction for the answer quality distribution under different categories of questions. However, there are some drawback for deep learning application on the ranking decision. First, it lacks a good interpretability because of the black box model, which could result in a controversy of users' experience. Second, it requires a large amount of data to train a big number of hyper-parameters, the local optima of limited dataset in many specific field might not do a good job for the ranking, especially for those fields in which the sample size is not as large enough to map the population distribution. Another problem is the cost and reliability of the training labels. Third, it is hard to steer the criterion if distinct answers deserve a high rank or not, which is though the strength of Wilson's method. In conclusion, among many factors in a Q&A community, the most direct and effective way for the high-quality answers ranking decision is to investigate the number of up-vote and down-vote, as Wilson score interval method does, which has a strong interpretability, fast deployability and easy adjustability online. This is the key motivation for improving Wilson score interval method.

## IMPROVED WILSON SCORE INTERVAL METHOD

A new modified method based on the Wilson Score Interval method was proposed in this paper. The equations are described as below:

$$\text{Improved Wilson Score Interval Method:}$$
$$(\text{with a fixed } z_{1-\frac{\alpha}{2}} \text{ and } P)$$
$$n = u + d$$
$$p = u/n$$
$$W(p, n) = \frac{p + \frac{z_{1-\frac{\alpha}{2}}^2}{2n} \pm \frac{z_{1-\frac{\alpha}{2}}}{2n}\sqrt{4n(1-p)p + z_{1-\frac{\alpha}{2}}^2}}{(1 + \frac{z_{1-\frac{\alpha}{2}}^2}{n})}$$
$$\text{find out } n_{max}$$
$$Score(u, n) = P \cdot W(p, n) + (1 - P) \cdot SI(u, n)$$

The first three equations are the same as original Wilson score interval method. $W(p, n)$ is equivalent to $W(u, n)$ since $p = u/n$. The last equation shows the final improved Wilson score is a weighted average of original Wilson score and the in-page Spotlight Index (SI in the equation). The Spotlight Index is defined as the degree or level of the current answer gaining attention with respect to the answer with the highest attention under the same question. $P$ is the weight of the Wilson score interval. A brief format for the Spotlight Index could be:

$$SI(u, n) = \frac{n}{n_{max}},$$

where $n_{max}$ is the voting number (including up-vote and down-vote) of the highest voting answer among all of the answers to the same question. For instance, if there are 3 answers in total to a question with respective voting numbers: 1, 50 and 100, their Spotlight Index are respectively 1/100, 50/100 and 100/100. Obviously, the range of Spotlight Index is [0, 1]. An answer will have higher Spotlight Index approaching to 1 when receiving more voting, and will have lower Spotlight Index approaching to 0 when receiving less voting. In order to avoid zero in the denominator, $n_{max}$ can be set to be 1 or a specific positive number at the very beginning. For instance, $n_{max}$ can be artificially defined as $n_{max} = 1$, if $n_{max}$ is found to be 0. Besides, The above equation could be also re-written as: $SI(u, n) = \frac{n}{n_{max}+1}$ or $SI(u, n) = \frac{n+1}{n_{max}+1}$. Please note that, $u$ does not appear in the right side of the equation since this equation only refers to the whole Spotlight Index definition and more definitions for the Spotlight Index will be given in the following table. $SI(u, n)$ can be written as $SI(u, n, n_{max})$, if $n_{max}$ is considered as a variable or parameter. In fact, $n_{max}$ is time dependent in most questions, which reflects how much attention the highest voted answer gains. Therefore, $n_{max}$ always changes with time, and is an implicit function of time. The introduction of $n_{max}$ makes the effect of the factor of time more accurate, meaningful and interpretable rather than the assumption of an explicit function form of time in answer quality ranking score. In some cases, $n_{max}$ could be manually manipulated to be less influential on the Spotlight Index weight at the very beginning of voting besides adjusting the more global parameter $P$. For example, let $n_{max}$ = a positive number (e.g. 10) if the actual $n_{max} < 10$. This is because there might be a larger bias when $n_{max}$ is small, and it is a convenient way to shrink the bias influence.

Since the Spotlight Index indicates how much attention does each answer under the same question gain, no matter if the attention is positive (more up-vote) or negative (more down-vote). This concept could be extended to a Spotlight Index series, as the table shows below:

*Table 1. Spotlight Index (SI) Series*

| | |
|---|---|
| (Whole) Spotlight Index | $\frac{u+d}{n_{max}}$ or $\frac{n}{n_{max}}$ |
| Net Spotlight Index | $\frac{u-d}{n_{max}}$ |
| Positive Spotlight Index | $\frac{u}{n_{max}}$ |
| Negative Spotlight Index | $-\frac{d}{n_{max}}$ |
| Up-vote Index | $\frac{u}{u_{max}}$ |
| Down-vote Index | $-\frac{d}{d_{max}}$ |

**First Draft was Completed in April 2018**

As the table above shows, the Spotlight Index series is divided into 6 categories. The definitions of $u$, $d$ and $n$ are as the same as those in original Wilson score interval method, and $n = u + d$. Besides, $u_{max}$ is the up-vote number of the highest up-voting answer among all of the answers to the same question and $d_{max}$ is the down-vote number of the highest down-voting answer among all of the answers to the same question. They are similar as $n_{max}$, but only take the number of up-vote or the number of down-vote into account respectively. Please note that $u_{max}$ and $d_{max}$ do not have to be from one same answer. In many questions, they belong to two different answers respectively. Therefore, in most cases, $n_{max} \neq u_{max} + d_{max}$. For the last two categories, Up-vote Index and Down-vote Index, the second last step of the improved Wilson score interval method "find out $n_{max}$" should be modified to be "find out $u_{max}$" and "find out $d_{max}$" respectively. Since Up-vote Index and Down-vote Index focus specifically on the up-vote or down-vote, they will not be mainly emphasized in this paper.

Besides the Spotlight Index series as above, there are other variants, such as logarithmic Spotlight Index series, exponential Spotlight Index series, polynomial Spotlight Index series, which are all nonlinear Spotlight Index series, compared to the linear series introduced above. The impact of different nonlinear-type Spotlight Index series on the variations of the improved Wilson's score are different. For instance, the logarithmic Spotlight Index series enable the improved Wilson's score to increase faster at the initial voting number and the increase slows down when the voting number becomes large. Let's take the logarithmic Whole Spotlight Index in Table 2 as an example, if during a certain time period, $n_{max} = 9999 \approx 10000$, and $\log(n_{max}) = 4$ (the base of the logarithm is set to be 10 by default in this paper), the Index requires only 9 votes from 0 increasing up to ¼, and requires additional 90 votes from ¼ increasing up to ½, and requires additional 900 votes from ½ increasing up to ¾, and requires additional 9000 votes from ¾ increasing up to 1. The logarithmic scale makes the increasing voting number 10 times harder to increase the Spotlight Index by the same amount. The hardness depends on the base of the logarithm. In an analogous manner, the exponential Spotlight Index series can be defined as to enable the improved Wilson's score to change slower at the initial votes and the speed of the change might vary (depends on the specific format of the exponential Spotlight Index series and the voting situation since the exponent part in some format could be negative) when the voting number becomes large, as Table 2 shows.

*Table 2. Some of Extended Spotlight Index (SI) Series*

| | |
|---|---|
| Logarithmic Whole Spotlight Index * | $\frac{\log(u+d+1)}{\log(n_{max}+1)}$ or $\frac{\log(n+1)}{\log(n_{max}+1)}$ |
| Logarithmic Net Spotlight Index * | $sgn(u-d) \cdot \frac{\log(|u-d|+1)}{\log(n_{max}+1)}$ |
| Logarithmic Positive Spotlight Index * | $\frac{\log(u+1)}{\log(n_{max}+1)}$ |
| Logarithmic Negative Spotlight Index * | $-\frac{\log(d+1)}{\log(n_{max}+1)}$ |
| Logarithmic Up-vote Index | $\frac{\log(u+1)}{\log(u_{max}+1)}$ |
| Logarithmic Down-vote Index | $-\frac{\log(d+1)}{\log(d_{max}+1)}$ |
| Exponential Whole Spotlight Index | $\frac{e^{u+d}}{e^{n_{max}}}$ or $\frac{e^n}{e^{n_{max}}}$ |
| Exponential Net Spotlight Index | $\frac{e^{u-d}}{e^{n_{max}}}$ |
| Exponential Positive Spotlight Index | $\frac{e^u}{e^{n_{max}}}$ |
| Exponential Negative Spotlight Index | $-\frac{e^d}{e^{n_{max}}}$ |
| Exponential Up-vote Index | $\frac{e^u}{e^{u_{max}}}$ |
| Exponential Down-vote Index | $-\frac{e^d}{e^{d_{max}}}$ |
| Polynomial Spotlight Index series † $(a > 0)$ | $\frac{n^a}{n_{max}{}^a}, \frac{sgn(u-d) \cdot (u-d)^a}{n_{max}{}^a}, \frac{u^a}{n_{max}{}^a},$ $-\frac{d^a}{n_{max}{}^a}, \frac{u^a}{u_{max}{}^a}, -\frac{d^a}{d_{max}{}^a}, \ldots\ldots$ |

*\* The base of logarithm is set to be 10 in this paper, and the extra adding 1 in the logarithm guarantees a non-negative number. Similar adding 1's operation can be applied to other Spotlight Index series such as polynomial. $sgn(u-d)$ is the sign function which determines if the sign of $u - d$ is positive or negative.*

*† The polynomial Spotlight Index series demonstrate a special case of it: the case of power function. The general polynomial case with lower order terms can be also defined and used in the same way. In this paper, we mainly focus on the power function case.*

In this section, Improved Wilson's method with different categories of the Spotlight Index will be compared with the original Wilson's method. The improved Wilson's method contains two parameters: $z_{1-\frac{\alpha}{2}}$ and $P$, which will be investigated for their impact on the final ranking score ($Score(u, n)$ in the equation) by the grid search algorithm. Please note that, if $z_{1-\frac{\alpha}{2}}$ and $P$ are not fixed as the equations above show, $W(p, n)$ and $Score(u, n)$ can be written as

**First Draft was Completed in April 2018**

$W(p, n, z_{1-\frac{\alpha}{2}})$ and $Score\left(u, n, P, z_{1-\frac{\alpha}{2}}\right)$. In the latter format, $z_{1-\frac{\alpha}{2}}$ and $P$ are also the parameters of the equations.

In order to grid search the impact of $z_{1-\frac{\alpha}{2}}$ and $P$, they are respectively assigned to be discrete values as below:
$$z_{1-\frac{\alpha}{2}} \in [0, 1, 5, 25], \quad P \in [0, 0.25, 0.5, 0.75, 1]$$
Different types of Spotlight Index in Table 1 will be also investigated to compare their roles played in the improved Wilson's score.

## RESULTS OF IMPROVED WILSON' METHOD

### THE OVERVIEW OF THE RESULTS

In this section, the improved Wilson score interval will be calculated with different parameters $z_{1-\frac{\alpha}{2}}$ and $P$ as discussed before and be visualized by contour plots. Please note that, the results will only show the lower bound of the improved Wilson score interval since the ranking decision for the answer quality usually concerns the lower bound as its criterion. But the upper bound can be investigated by using the same way. The range of $u$ and $d$ are both [0, 1000], without loss of generality. $n_{max}$ is set to be 2000 for covering all values of $u + d$. However, as we discussed, $n_{max}$ is a time dependent variable or parameter in the real world, and we will discuss the effect of different $n_{max}$ in the last part of this section.

First, Figure 1 and 2 show a 3D contour plots about the original and some versions of improved Wilson's score (the lower bound by default). Figure 1 shows how the improved Wilson scores differ from the original Wilson score, especially for the trend near the highest and lowest scores area. Figure 2 shows the non-zero $z_{1-\frac{\alpha}{2}}$ value (as Z=2 in the figure) makes the Wilson's correction from the traditional average rating method [Miller, 2009], especially considering the uncertainty when the number of votes is small. It is important to notice that the improved Wilson score inherits this correction from the original Wilson's method. Besides, the improved Wilson score with different Spotlight Index has their own strength and characteristics in answer quality ranking decision. More details about the difference will be discussed.

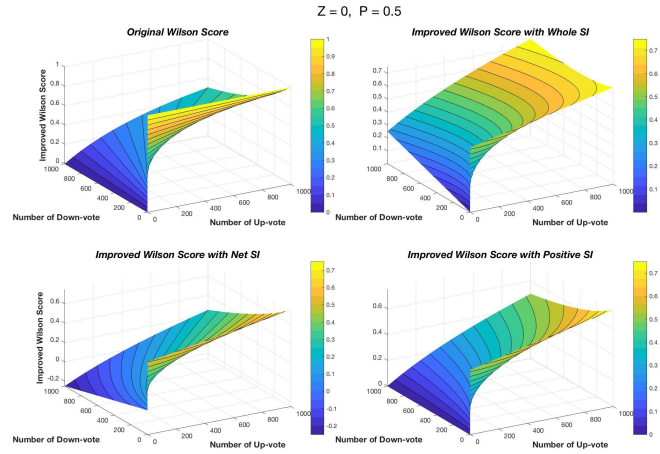

*Figure 1. The 3D contour plot of original Wilson Interval Score method and some improved Wilson's methods, with $z_{1-\frac{\alpha}{2}} = 0$ and $P = 0.5$.*

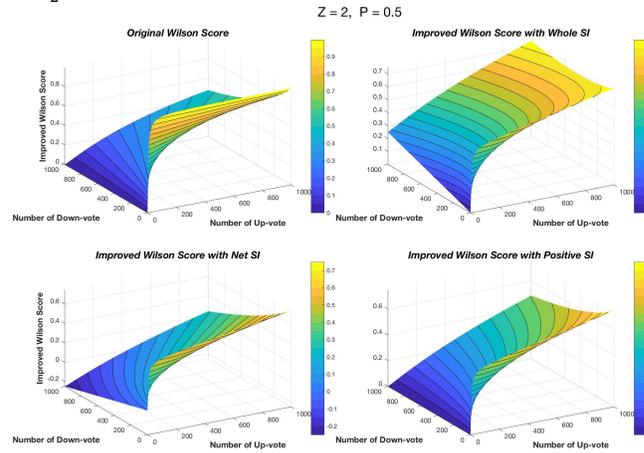

*Figure 2. The 3D contour plot of original Wilson Interval Score method and some improved Wilson's methods, with $z_{1-\frac{\alpha}{2}} = 2$ and $P = 0.5$.*

In order to better compare the difference between the original Wilson's methods and improved Wilson's method with different Spotlight Index. Some of the results are shown in 2D contour plot in Figure 2 – Figure 6. Each of these figures contains the results of original Wilson score, improved Wilson score with Whole Spotlight Index, Net Spotlight Index and Positive Spotlight Index in order, as the figures show. Based on the equations discussed, $P$ value will not affect the original Wilson score but the improved ones.

### THE INFLUENCE OF P VALUE

Figure 3 shows the case when $P = 0$, the original Wilson score across the number of up-vote and down-vote from 0 to 1000, as we familiar to the Wilson score interval equation. The improved Wilson score with different Spotlight Index is shrunk into their respective Spotlight Index function only, as $P = 0$ discards the Wilson score part. The Whole SI receives highest score in the upper right area and lowest in the lower left area. Compared, the Net SI receives highest score in the

**First Draft was Completed in April 2018**

lower right area and lowest in the upper left area. The Positive SI receives the highest score in the right side and lowest score in the left side. As we expect, the focus of each version of SI

Wilson score with different SI, comparing to Figure 3. Therefore, in this case, the Spotlight Index dominates the total score for answer ranking.

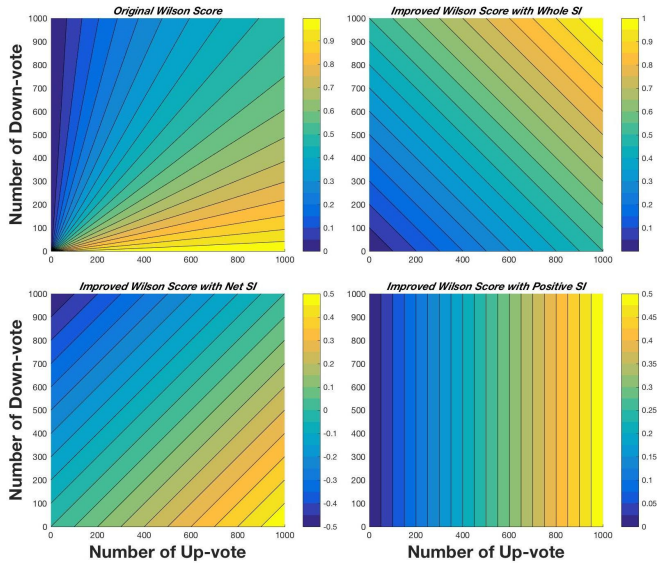

*Figure 3. The 2D contour plot of original Wilson Interval Score method and some improved Wilson's methods, with $z_{1-\frac{\alpha}{2}} = 2$ and $P = 0$.*

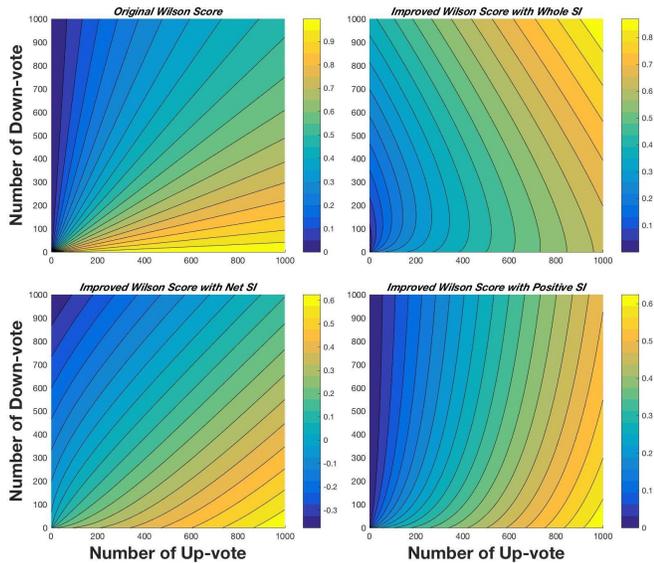

*Figure 4. The 2D contour plot of original Wilson Interval Score method and some improved Wilson's methods, with $z_{1-\frac{\alpha}{2}} = 2$ and $P = 0.25$.*

was consistent with their definition in Table 1. And their specialities would correct the original Wilson score with different focuses as needed by a certain amount, which is determined by the value of $P$.

Figure 4 reveals the case when $P = 0.25$. In other words, the weight of the SI part out of the total improved Wilson score is three times of that of the original Wilson score part. The original Wilson score marginally modified the improved

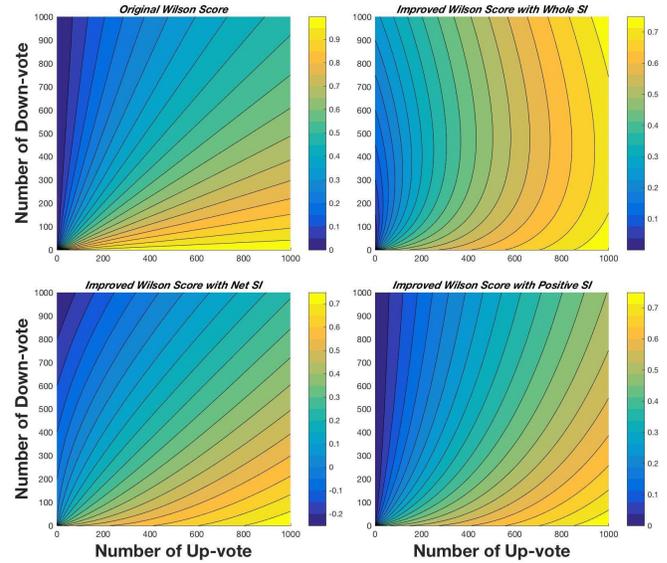

*Figure 5. The 2D contour plot of original Wilson Interval Score method and some improved Wilson's methods, with $z_{1-\frac{\alpha}{2}} = 2$ and $P = 0.5$.*

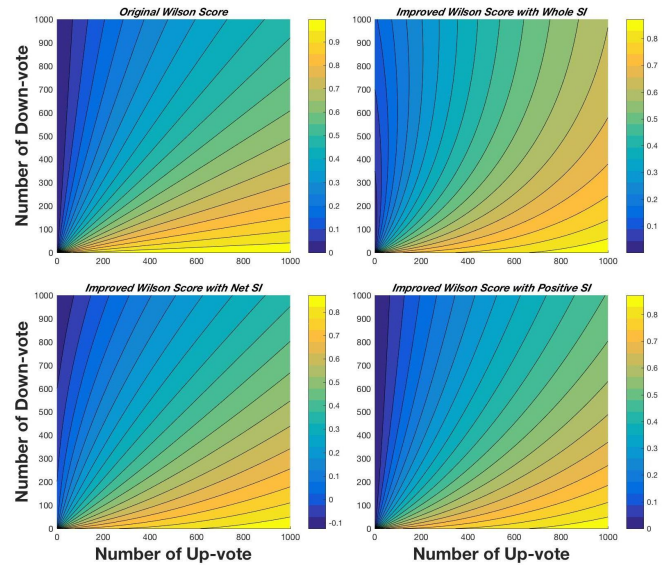

*Figure 6. The 2D contour plot of original Wilson Interval Score method and some improved Wilson's methods, with $z_{1-\frac{\alpha}{2}} = 2$ and $P = 0.75$.*

Figure 5 reveals the case when $P = 0.5$. The weights were equally distributed into the original Wilson score term and the Spotlight Index term. It is an effective strategy which can be frequently applied in practice where the Spotlight Index is considered as of equal importance as the original Wilson score. As the figure shows, the improved Wilson score with Whole Spotlight Index focuses not only on the ratio of

**First Draft was Completed in April 2018**

up-vote number to the total number of votes, but also on the total number of votes for each answer in the vote space. It takes the general attention of all voters into account by introducing the Whole Spotlight Index. The contour in the plot changes from the radially straight lines as the original Wilson score shows to the curved lines bent towards the upper right area. The improved Wilson score with Net SI considered both attention from up-votes and down-votes, calculating how much pure up-votes attention by subtracting the number of down-votes received in each answer. The strength of Net SI is that it allows degradation of the total ranking score if an answer is considered of low quality by the majority of voters even in the very beginning, so that the low-quality answer will be ranked to fall behind the new, no rated or few rated answers. However, the pure up-vote attention mechanism might dim to distinct the ranking of the answers with nearly same amount of up-votes and down-votes, since their Net Spotlight Indices approach to zero. The Net Spotlight Index makes the contour lines bent towards the lower right area if it is positive and towards the upper left area if it is negative.

Figure 6 reveals the case when $P = 0.75$. In other words, the weight of the original Wilson score part out of the total improved Wilson score is three times of that of the SI part. In this case, the original Wilson score dominates the total score for the answer quality ranking. Therefore, the original Wilson score is slightly modified by the different Spotlight Indices discussed before. If a community has been using the original Wilson score interval method for the answer quality ranking, it is recommended to use such a conservative $P$ value (or even a larger value, e.g., 0.9) to smoothly transfer from the original Wilson score strategy to an improved Wilson score strategy, without largely affecting or shapely changing the experience of users.

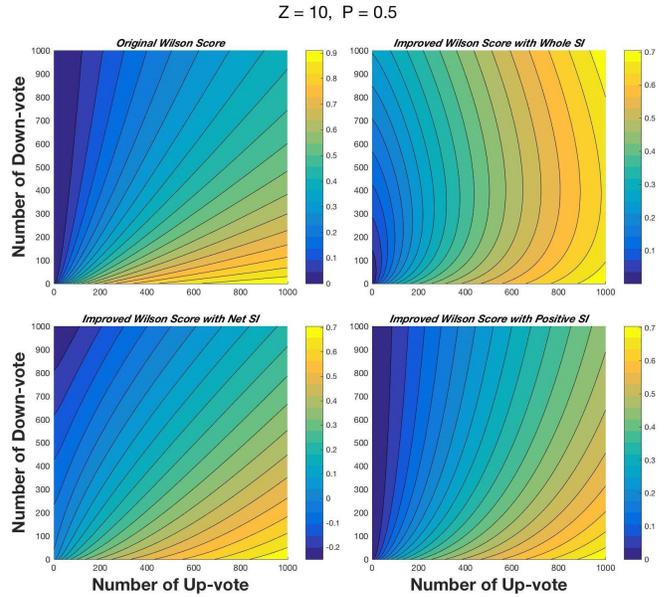

*Figure 8. The 2D contour plot of original Wilson Interval Score method and some improved Wilson's methods, with $z_{1-\frac{\alpha}{2}} = 10$ and $P = 0.5$.*

### THE INFLUENCE OF $z_{1-\alpha/2}$ VALUE

For the sake of comparing the effect of $z_{1-\frac{\alpha}{2}}$, the results of $z_{1-\frac{\alpha}{2}} = 5$ and $z_{1-\frac{\alpha}{2}} = 10$ with $P = 0.5$ are demonstrated in the Figure 7 and 8. The increasing $z_{1-\frac{\alpha}{2}}$ makes the density of the contour sparser and the highest scores are distributed more concentrated in the bottom area. Therefore, the parameter $z_{1-\frac{\alpha}{2}}$ is mainly responsible for modifying the average rating method, combined with the parameter $P$ which is responsible for the influence of the voters' attention. They together modulate the rule of scoring for answer quality ranking. Please note that, since the range of Net SI is $[-1, 1]$, the improved Wilson score with it is $P \cdot [0, 1] + (1 - P) \cdot [-1, 1]$, which is $[P - 1, 1]$. Therefore, the improved Wilson score with Net SI could be negative if $P \neq 1$. And the negative score indicates that the answer is of low quality even compared to the non-rated answers. It is validated that the Negative SI (not shown here) could also make the improved Wilson score negative values, which can be used to degrade low-quality answers.

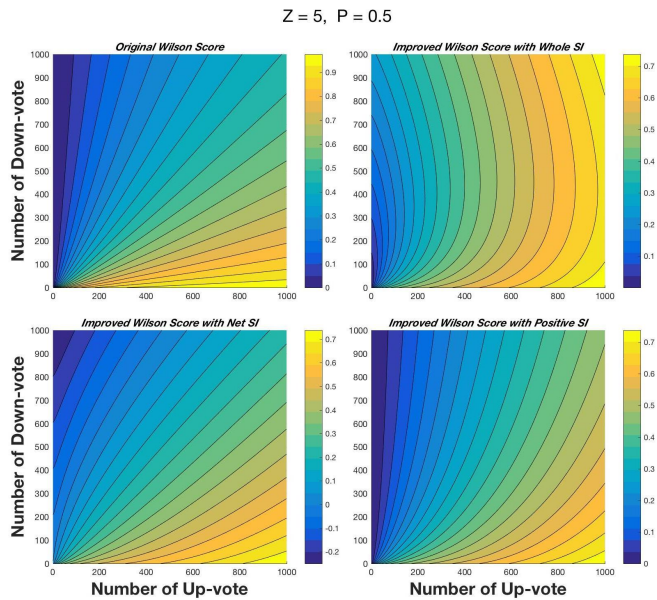

*Figure 7. The 2D contour plot of original Wilson Interval Score method and some improved Wilson's methods, with $z_{1-\frac{\alpha}{2}} = 5$ and $P = 0.5$.*

**First Draft was Completed in April 2018**

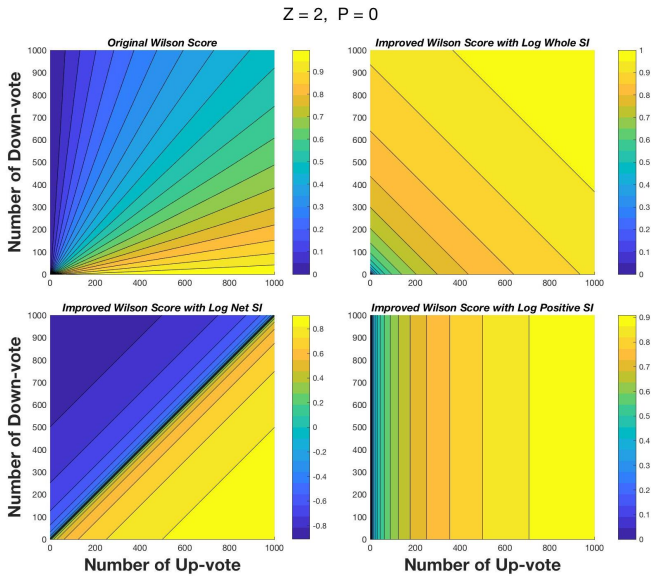

*Figure 9. The 2D contour plot of original Wilson Interval Score method and some improved Wilson's methods with logarithmic Spotlight Indices, with $z_{1-\frac{\alpha}{2}} = 2$ and $P = 0$.*

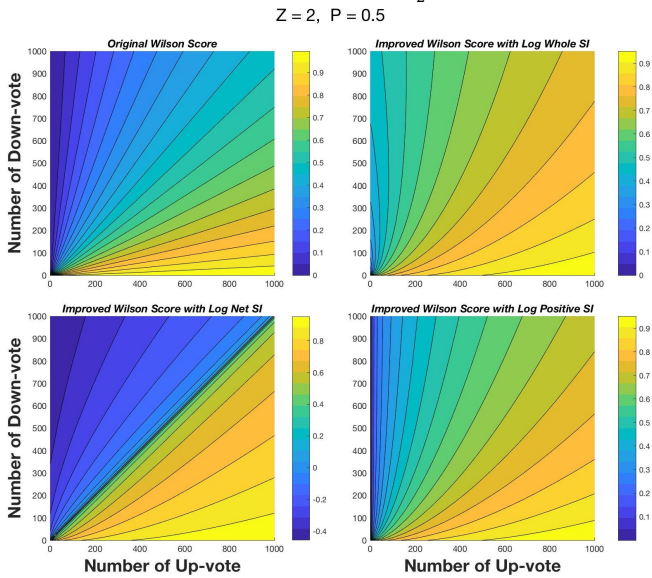

*Figure 10. The 2D contour plot of original Wilson Interval Score method and some improved Wilson's methods with logarithmic Spotlight Indices, with $z_{1-\frac{\alpha}{2}} = 2$ and $P = 0.5$.*

Figure 9 and 10 demonstrate the results of the improved Wilson score with some of the logarithmic Spotlight Indices. Compared with the "linear" Spotlight Index cases, the logarithmic Spotlight Indices make the contour denser in the area where the SI values close to zero value, and sparser in the area where the SI values away from zero value. This feature allows the scores change faster in the beginning of the vote, except for a special case of Net SI when the number of up-vote and down-vote always remain close to each other.

The gradient of exponential Spotlight Index is very high with the variation of the voting number, it usually just brings in a small number compared to the original Wilson score. Instead, the polynomial Spotlight Index performs a nearly opposite feature of the logarithmic Spotlight Index, as Figure 11 shows.

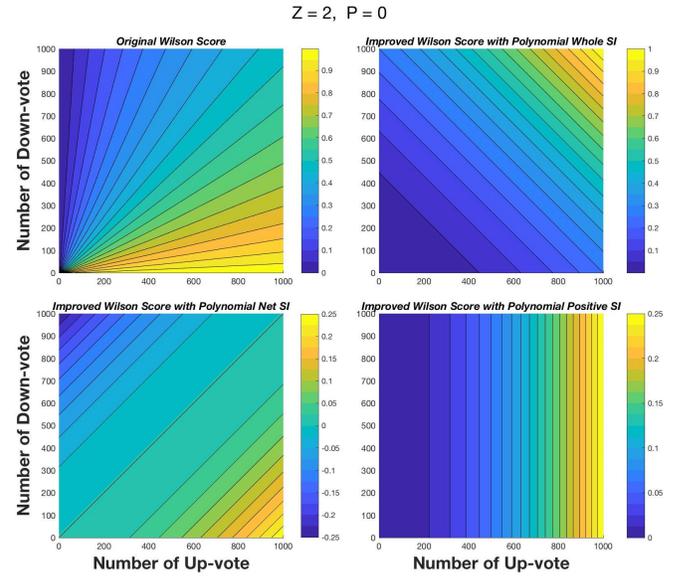

*Figure 11. The 2D contour plot of original Wilson Interval Score method and some improved Wilson's methods with polynomial Spotlight Indices (the exponent $a = 2$), with $z_{1-\frac{\alpha}{2}} = 2$ and $P = 0$.*

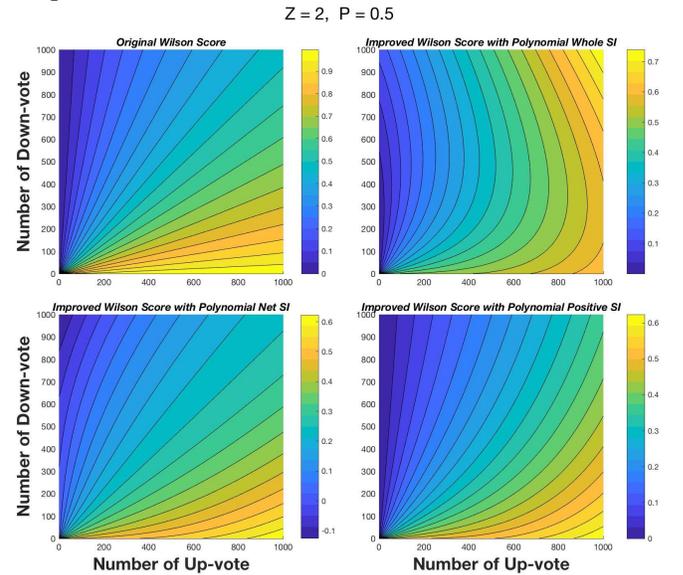

*Figure 12. The 2D contour plot of original Wilson Interval Score method and some improved Wilson's methods with polynomial Spotlight Indices (the exponent $a = 2$), with $z_{1-\frac{\alpha}{2}} = 2$ and $P = 0.5$.*

Figure 12 shows the improved Wilson score with polynomial Spotlight Indices. Compared with the "linear" Spotlight Index cases, the polynomial Spotlight Indices make the contour sparser in the area where the SI values close to zero value, and denser in the area where the SI values away from zero value. This feature allows the scores change slower in the beginning of the vote, except for a special case of Net SI when the

**First Draft was Completed in April 2018**

number of up-vote and down-vote always remain close to each other, but change faster when the voting number increases.

In theory, we can use the Spotlight Index of general polynomial functions combined with a proper *P* value to regulate the speed of the improved Wilson score's variation in different periods of voting, instead of a homogeneous feature as that of the "linear" Spotlight Indices as earlier discussed.

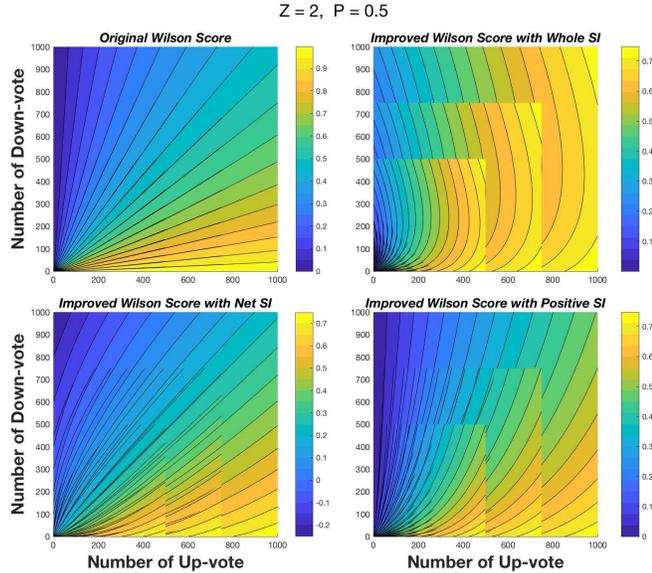

*Figure 13. The 2D contour plots of original Wilson Interval Score method and some improved Wilson's methods with $z_{1-\frac{\alpha}{2}} = 2$ and $P = 0.5$. Every subfigure contains three superimposed results, which show the scores with the number of up-vote and down-vote in the range of [0, 500], [0, 750], [0, 1000].*

Figure 13 reveals the results similar as Figure 5, but is superimposed by the two other results which are calculated by assuming that the current $n_{max} = 1500$ or $n_{max} = 1000$ instead of $n_{max} = 2000$. In other words, the maximum number of up-vote and down-vote are both 750 or both 500. As we discussed, if considering $SI(u,n)$ as $SI(u,n,n_{max})$, the improved Wilson score is also a function of $n_{max}$. In fact, $n_{max}$ usually keeps changing with the dynamic voting process, which could be considered an implicit function of time. The results show that no matter how $n_{max}$ changes, the pattern or profile of the improved Wilson score adaptively changes with the consistent scaling. This feature guarantees the scores proportionally adapt for the overall voting size.

## SUMMARY

In conclusion, we proposed an improved Wilson Score Interval method for the community answer quality ranking problem. We introduced the Spotlight Index term series, which successfully modified the original Wilson Interval Score method by different focuses. We investigated the influences of P values and $z_{1-\frac{\alpha}{2}}$ values in the equation of the improved method, which revealed a good performance on diverse and complex application scenarios with a strong interpretability about the voters' attention. The improved method could be time dependent but remains computationally economical as the original Wilson Score Interval method. The improved method could be widely used in the answer quality ranking communities, such as Reddit, Zhihu, and in the voting section for the customer reviews (e.g. vote for Helpful or Not Helpful) for the electronic commerce companies such as Amazon, eBay, JingDong, TaoBao, or the up-vote or down-vote for the video comments in the YouTube, Hulu, which can help to improve the answer quality ranking, the shopping experience of the customers and the watching experience of the audiences.

## AUTHOR INFORMATION

Xin Cao, PhD candidate in Computational Space Physics & Master of Computational Science and Engineering, Georgia Institute of Technology.

**First Draft was Completed in April 2018**